\documentclass[sigconf]{acmart}
\AtBeginDocument{%
  \providecommand\BibTeX{{%
    \normalfont B\kern-0.5em{\scshape i\kern-0.25em b}\kern-0.8em\TeX}}}

\copyrightyear{2022} 
\acmYear{2022} 
\setcopyright{acmlicensed}
\acmConference[ITiCSE 2022]{Proceedings of the
27th ACM Conference on Innovation and Technology in Computer Science
Education Vol 1}{July 8--13, 2022}{Dublin, Ireland}
\acmBooktitle{Proceedings of the 27th ACM Conference on Innovation and
Technology in Computer Science Education Vol 1 (ITiCSE 2022), July 8--13,
2022, Dublin, Ireland}
\acmPrice{15.00}
\acmDOI{10.1145/3502718.3524772}
\acmISBN{978-1-4503-9201-3/22/07}

\newcommand{\fix}[1]{\textcolor{red}{\textbf{\textit{#1}}}}

\begin{document}

\fancyhead{}

\title{The Impact of Remote Pair Programming in an Upper-Level CS Course}

\author{Zachariah Beasley}
\email{zjb@usf.edu}
\orcid{0000-0002-0146-2739}
\affiliation{%
  \institution{University of South Florida}
  \streetaddress{4202 E. Fowler Ave.}
  \city{Tampa}
  \state{Florida}
  \country{USA}
  \postcode{33620}
}
\author{Ayesha Johnson}
\email{arjohns2@usf.edu}
\orcid{0000-0003-2632-9157}
\affiliation{%
  \institution{University of South Florida}
  \streetaddress{4202 E. Fowler Ave.}
  \city{Tampa}
  \state{Florida}
  \country{USA}
  \postcode{33620}
}

\renewcommand{\shortauthors}{Beasley and Johnson}

\begin{abstract}
  Pair programming has been highlighted as an active learning technique with several benefits to students, including increasing participation and improving outcomes, particularly for female computer science students. However, most of the literature highlights the effects of pair programming in introductory courses, where students have varied levels of prior programming experience and thus may experience related group issues. This work analyzes the effect of pair programming in an upper-level computer science course, where students have a more consistent background education, particularly in languages learned and best practices in coding. Secondly, the effect of remote pair programming on student outcomes is still an open question and one of increasing importance with the advent of Covid-19. This work utilized split sections with a control and treatment group in a large, public university. In addition to comparing pair programming to individual programming, results were analyzed by modality (remote vs. in person) and by gender, focusing on how pair programming benefits female computer science students in confidence, persistence in the major, and outcomes. We found that pair programming groups scored higher on assignments and exams, that remote pair programming groups performed as well as in person groups, and that female students increased their confidence in asking questions in class and scored 12\% higher in the course when utilizing pair programming.
\end{abstract}



\maketitle

\section{Introduction}
Pair programming, where two students share one screen while working on a programming assignment, has often been heralded as an effective active learning technique to increase participation and improve outcomes in the classroom, particularly for women in computer science. One student, the driver, writes the program, while the other, the navigator, provides direction, encouragement, and debugs in real-time. Student benefits from this collaborative learning approach include 1) the opportunity for ``continuous review" where defects are corrected as they arise, 2) greater satisfaction in the course and learning from peers, 3) team building, communication, and improvement of other soft skills, and 4) greater industry and productivity \cite{williams2001support,cockburn2000costs}. Researchers have discovered a 95\% increase in confidence in the final product and found that ``pair programming is 40-50\% faster than programming alone" \cite{mcdowell2002effects}. Others have noted that ``programming assignment grades, exam scores, and persistence in computer programming courses" improve if pair programming is done properly \cite{umapathy2017meta}. For instructors, pair programming can provide a barrier to academic dishonesty, promote healthy social interaction in a post-pandemic environment of relative isolation, and allow students to coach and learn from one another \cite{williams2001support}.

For all its benefits, there are still several open questions in pair programming. One is the effect of \textit{remote} pair programming on student outcomes, i.e., whether students must be physically together to experience the full benefits. This question is of particular importance post-pandemic as students are more comfortable with online work. As our university transitioned back to the in person modality, we found that most students (95\%) preferred to work on assignments remotely. A second consideration is the optimal pairing of programming partners. For example, Bowman et al. found a negative effect when pairing students with significantly different programming ability \cite{bowman2019prior}. Since most of the literature highlights the effect of pair programming in introductory courses, partners can have a wide divergence in ability and thus experience problems with one partner leaving the other behind. Relatively little work has studied pair programming in upper-level computer science courses where students have a more consistent background education. Thus, the goals of this work are to examine the effect of pair programming 1) where students could work remotely on assignments and 2) on an \textit{upper-level}, CS-3 course where students should exhibit greater consistency in education, style, and languages learned. This work utilized split sections with a control and treatment group at a large, public university. Results were analyzed with a particular focus on how pair programming benefits female computer science students in confidence, persistence in the major, and outcomes.

\section{Previous Research}

Much of the prior work on pair programming in computer science education has studied its effects in introductory courses, where most students have little or no formal exposure to programming languages and paradigms. Furthermore, those students who do have prior experience come from a variety of backgrounds with various levels of competency and languages learned. Students in a pair programming group in such an introductory course can experience mixed results, often related to the differing levels of prior experience. For example, one large scale study conducted over two years in three different introductory courses (CS, information science, and humanities) at the University of Iowa found that ``students who worked with a more experienced partner actually had poorer outcomes, including lower effort exerted on the assignment, perceptions that their partner gave more effort than they did, less time in the driving role (i.e., typing out the assignment), lower understanding of concepts from lab, and less interest in computer science overall" \cite{bowman2019prior}. Another study found that women in an introductory CS course who were randomly paired on assignments found it beneficial to have ``someone to ask questions and discuss ideas [with]", especially before they would approach a TA, and that it ``improves understanding" \cite{ying2019their}. But several women in the same study also reported a feeling that they were ``burdening their partner" due to less programming experience. Finally, the unknowns in how to best assign pair programming partners and how to navigate complex partner interaction has led some to propose automating the pair programming partner as a conversational agent that adapts to the user's skill level \cite{robe2021designing}. While an exciting research direction, the ability of such an agent to solve complex problems as either the driver or the navigator is currently infeasible.

In regards to remote pair programming, a study of distributed \textit{team programming} (with two to four students per team) in an upper level CS course between two universities, University of North Carolina and North Carolina State University, found no statistically significant productivity or quality differences between geographically distributed programming teams and in person teams \cite{baheti2002exploring}. A more recent work has studied the effects of leadership style and ``pair jelling" in a remote and part-time educational setting, with a class of size 10, but did not investigate remote pair programming's effect on grades or student outcomes \cite{adeliyi2021investigating}. Several recent small-scale studies have weighed in on remote pair programming. One, limited to six pairs, found that women use more non-verbal cues and prefer co-located over remote pair programming to feel connected \cite{kuttal2019remote}. In another, with four participants paired by a research team according to perceived confidence and experience, remote pair programming was ``associated positively with feelings
about communication skills and team-working, which
may enhance employability" \cite{hughes2020remote}. However, no quantitative work on the effects of pair programming was performed. In general, it has been suggested that much work remains to establish the effectiveness of remote pair programming on student outcomes, especially since it is increasing in utilization in both academia and industry \cite{da2015distributed}.

\section{Course Description}
Data Structures is a 400-level course offered in the Department of Computer Science and Engineering at a large, public university. The course textbook was Data Structures and Algorithms in C++ (2nd Edition) by Goodrich, Tamassia, and Mount. Course materials were taught primarily through live coding, lectures, and small group discussion.  Students received 150 minutes of instruction per week, with access to the professor, teaching assistants, and peer leaders \cite{clark2020peer} outside of contact hours. Data Structures was offered in person, with code and lecture recordings available for those who were unable to attend due to quarantine or other health-related reasons. 

During the 2021 fall semester two identical sections of Data Structures were taught by the first author, one of which utilized pair programming (section 1) and a control group which did not (section 2). The pair programming section had 62 students (20 of whom were women) who were randomly assigned the same pair for all four graded assignments. The second section had 45 students (4 of whom were women) who completed each assignment individually.

During the second week of the semester, before any programming projects were assigned, the instructor taught object-oriented design concepts through a week-long live coding project. In this way, a form of pair programming was demonstrated to students, with the instructor as the driver and the class collectively as the navigator. In pair programming, the driver types the code while the navigator develops or modifies the flow of logic, suggests corrective, perfective, or refactorative maintenance, and encourages. Both the driver and the navigator participate real-time by watching the same screen, in person or remotely via screen sharing. Prior to beginning the first assignment, students in section 1 were instructed to read a document on pair programming best practices \cite{williams2000all}.

\section{Methods}

\subsection{Research Design and Participants}
Participants included all Data Structures students. Data was collected from student records, including student examination scores and assignment completion.
In addition, we conducted a pre- and post-course survey asking students to self-evaluate the effectiveness of active learning practices (pair programming in detail) and to report on confidence, persistence in the major, and preference. The 4-minute surveys were voluntary and were sent by another instructor to mitigate perceived coercion. The survey instruments were piloted before the semester. Each survey had an 8-day window for completion, with students receiving extra credit on the final exam for successful completion of both surveys. Alternative means of obtaining equivalent extra credit was provided. The pre-course survey had a completion rate of 87\% and the post-course survey had a completion rate of 92\%.

\subsection{Measures}
Students completed the first survey (described above) at the beginning of the semester to gather demographic information and capture attitudes toward computer science. At the end of the semester students completed the second survey which included additional questions related to their experience in the course. The pair programming section was also asked questions related to their experiences using pair programming.

Students were given 4 programming assignments (weighted 40\%) and 3 exams (weighted 60\%) throughout the semester, with other practice assignments not counting toward final course grade. All exams were completed individually. A final score for the class was given based on a composite score of all course work and exams.

\subsection{Data Analysis}
The primary outcome explored was students’ final score in the class. This score was compared between the pair programming section and the section working individually using an independent samples t-test. Separate repeated measures ANOVA models were estimated to explore exam scores over time, and scores on assignments over time across the two groups. Differences found were further explored using post hoc mean comparisons. The type I error rate in post hoc analyses were adjusted using Bonferroni’s adjustment. Additionally, gender differences and differences between pairs working remotely versus in person were explored.

\section{Results}

A total of 88 students completed both surveys: 64 male students and 24 female students. Pair programming (n=56) and individual sections (n=32) were similar at both at baseline and end of the term with respect to responses given on the survey with the exception of gender. There were more women in the pair programming section than in the individual programming section (20 versus 4). 

\subsection{Individual vs. Pair Programming}
\begin{figure}[h]
  \centering
  \includegraphics[width=\linewidth]{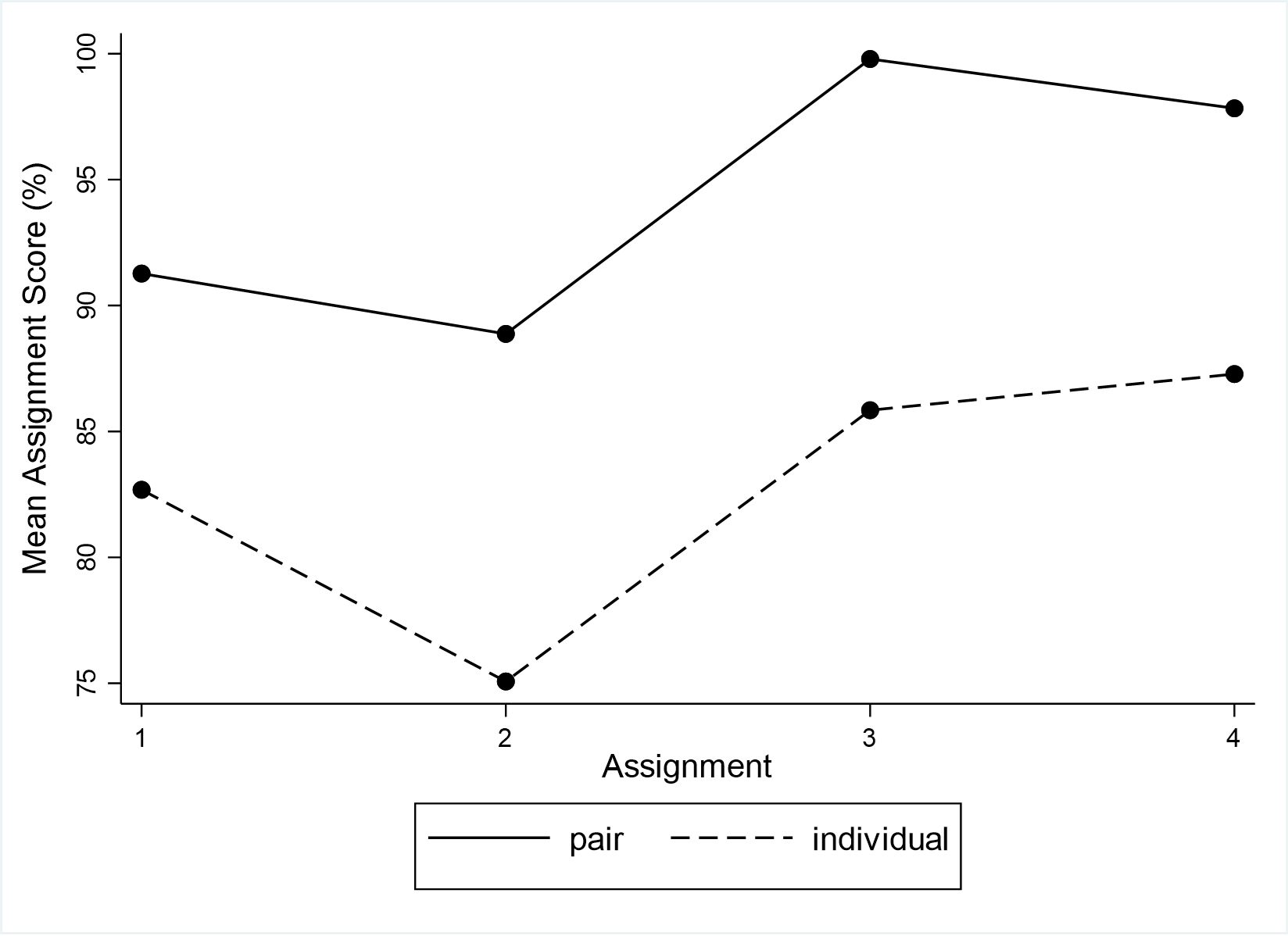}
  \caption{Pair versus Individual Assignment Scores}
  \label{fig:assignmentScore}
\end{figure}

In \autoref{tab:ppOnScores}, repeated measures ANOVA on programming assignments showed both a within-subject (p = .003) and a between subject effect (p = .001). The pair programming group consistently scored higher than the individuals over time, with the difference ranging from 8--14\% (\autoref{fig:assignmentScore}). Additionally, the pair programming group scored nearly perfectly on assignments 3 and 4.

\begin{figure}[h]
  \centering
  \includegraphics[width=\linewidth]{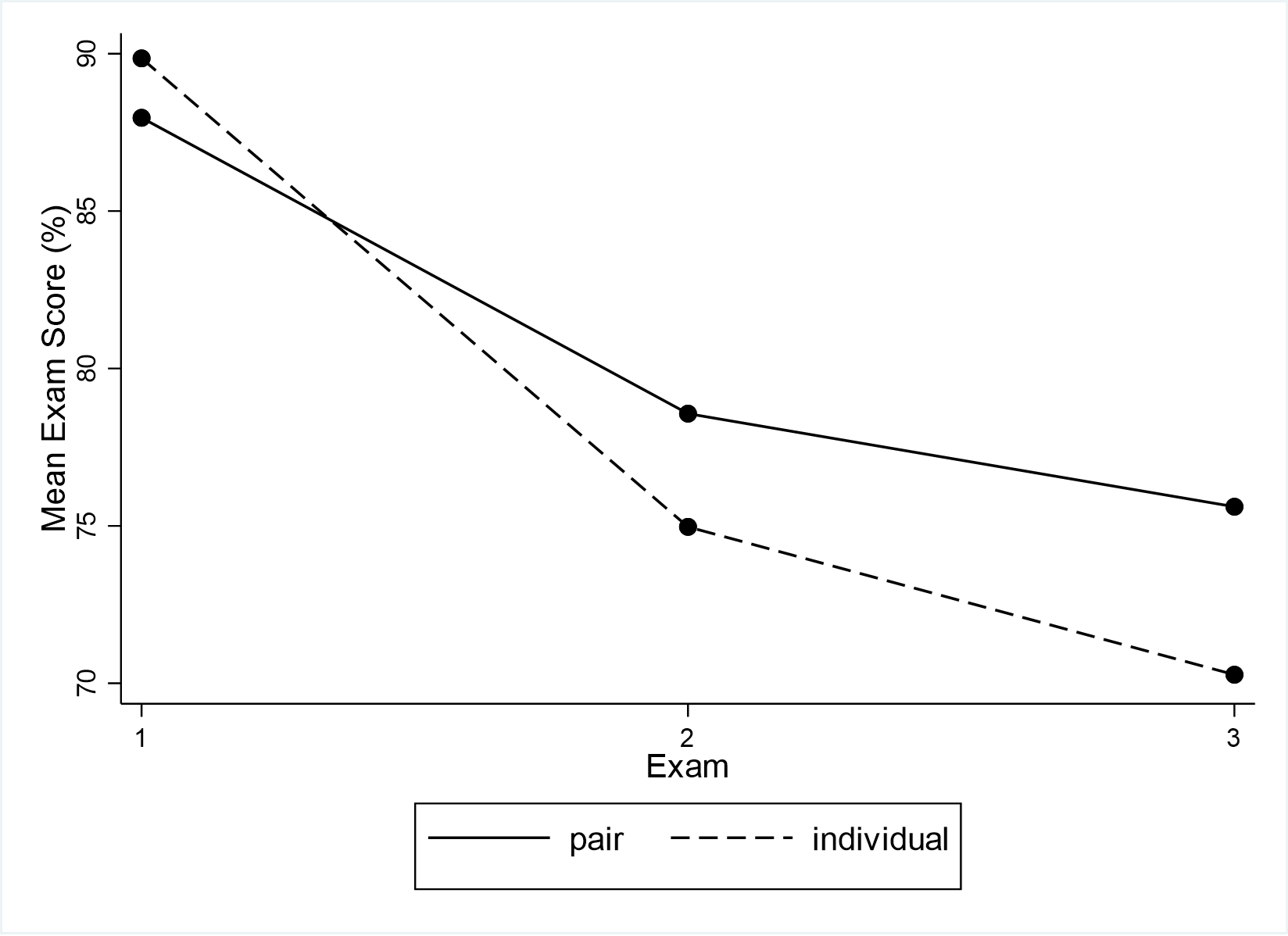}
  \caption{Pair versus Individual Exam Scores}
  \label{fig:examScore}
\end{figure}

Repeated measures ANOVA on exam score showed a within-subject effect (p < .001) but no between subject effect. The pair programming group scored higher than the individual programming group as the semester progressed (\autoref{fig:examScore}). Final scores revealed a 7\% higher score among students in the pair programming group (87.4 vs. 80.4; p = .013) compared to students in the individual programming group (\autoref{tab:ppOnScores}).

\begin{table*}[h]
  \caption{Effect of Pair Programming on Scores}
  \label{tab:ppOnScores}
      \scalebox{0.95}{
\begin{tabular}{lrrllr}
\toprule
Section&Pair&Individual&Test&Effect&P-value\\
&Mean (SD)&Mean (SD)&&&\\
\midrule
Assignment 1&91.3 (19.3)&82.7 (32.1)&RM-ANOVA&Group&0.001\\
Assignment 2&88.9 (27.6)&75.1 (34.9)&&Time&0.003\\
Assignment 3&99.8 (1.1)&85.8 (31.6)&&Group-Time&0.733\\
Assignment 4&97.8 (13.0)&87.3 (32.5)&&&\\
&&&&&\\
Exam 1&88.2 (12.3)&90.0 (8.7)&RM-ANOVA&Group&0.37\\
Exam 2&78.1 (23.0)&74.2 (20.9)&&Time&<.001\\
Exam 3&75.6 (17.4)&70.3 (22.6)&&Group-Time&0.306\\
&&&&&\\
Final Score&87.4 (9.7)&80.4 (16.8)&T-test&Group&0.013\\
Final Score (Women)&83.0 (10.4)&71.4 (21.0)&T-test&Group&0.091\\
  \bottomrule
\end{tabular}
}
\end{table*}

\begin{figure}[h]
  \centering
  \includegraphics[width=\linewidth]{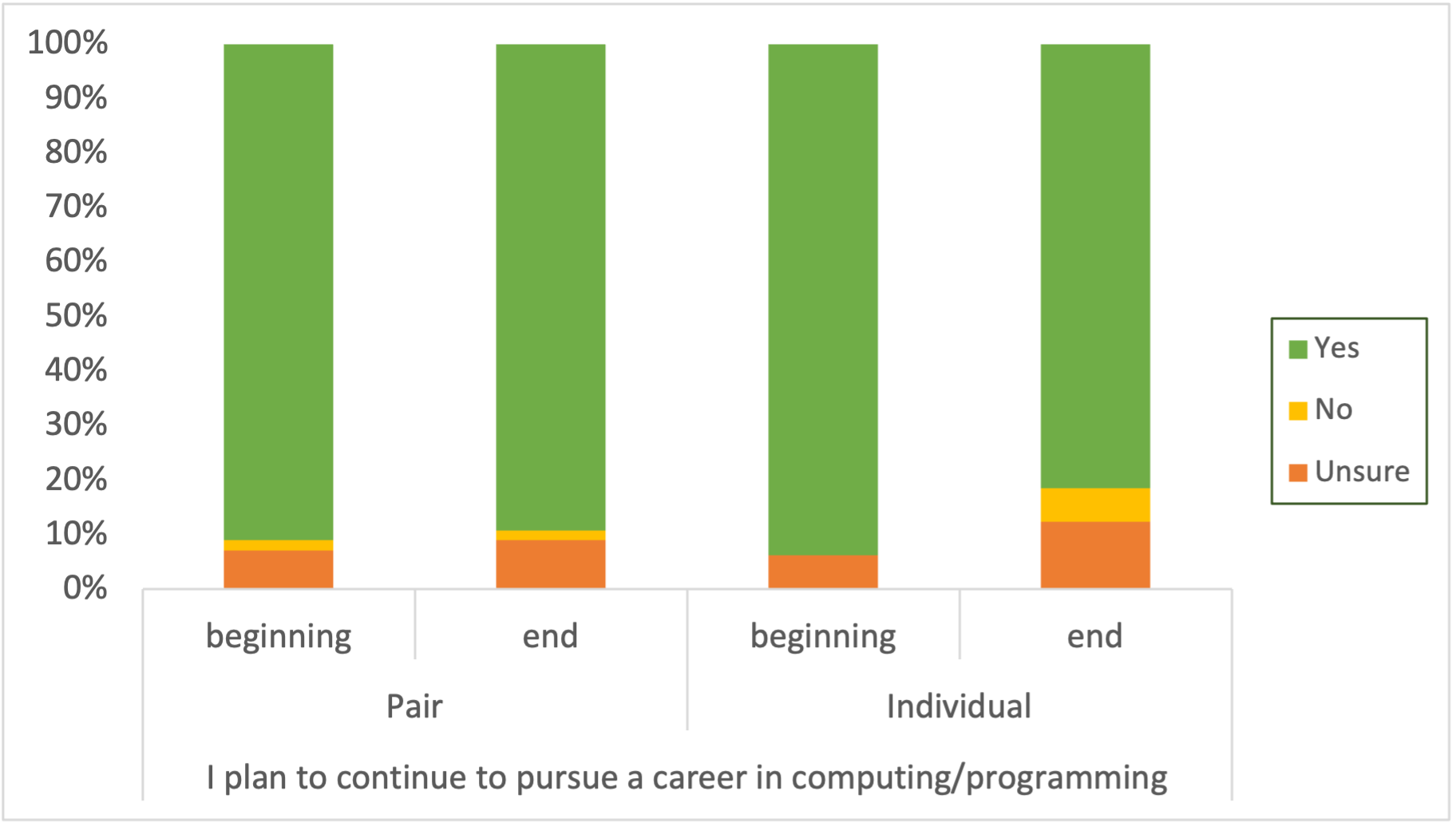}
  \caption{Pair versus Individual Plan to Continue in Computing}
  \label{fig:pCareer}
\end{figure}

From the survey, for both paired and individual students, there was a slight decrease in the number of students planning ``to continue to pursue a career in computing/programming" (\autoref{fig:pCareer}). This effect was mitigated in the pair programming group, which saw a total increase of 1.8\% in those selecting ``no" or ``unsure", as opposed to an increase of 12.3\% in the individual group.

\begin{figure}[h]
  \centering
  \includegraphics[width=\linewidth]{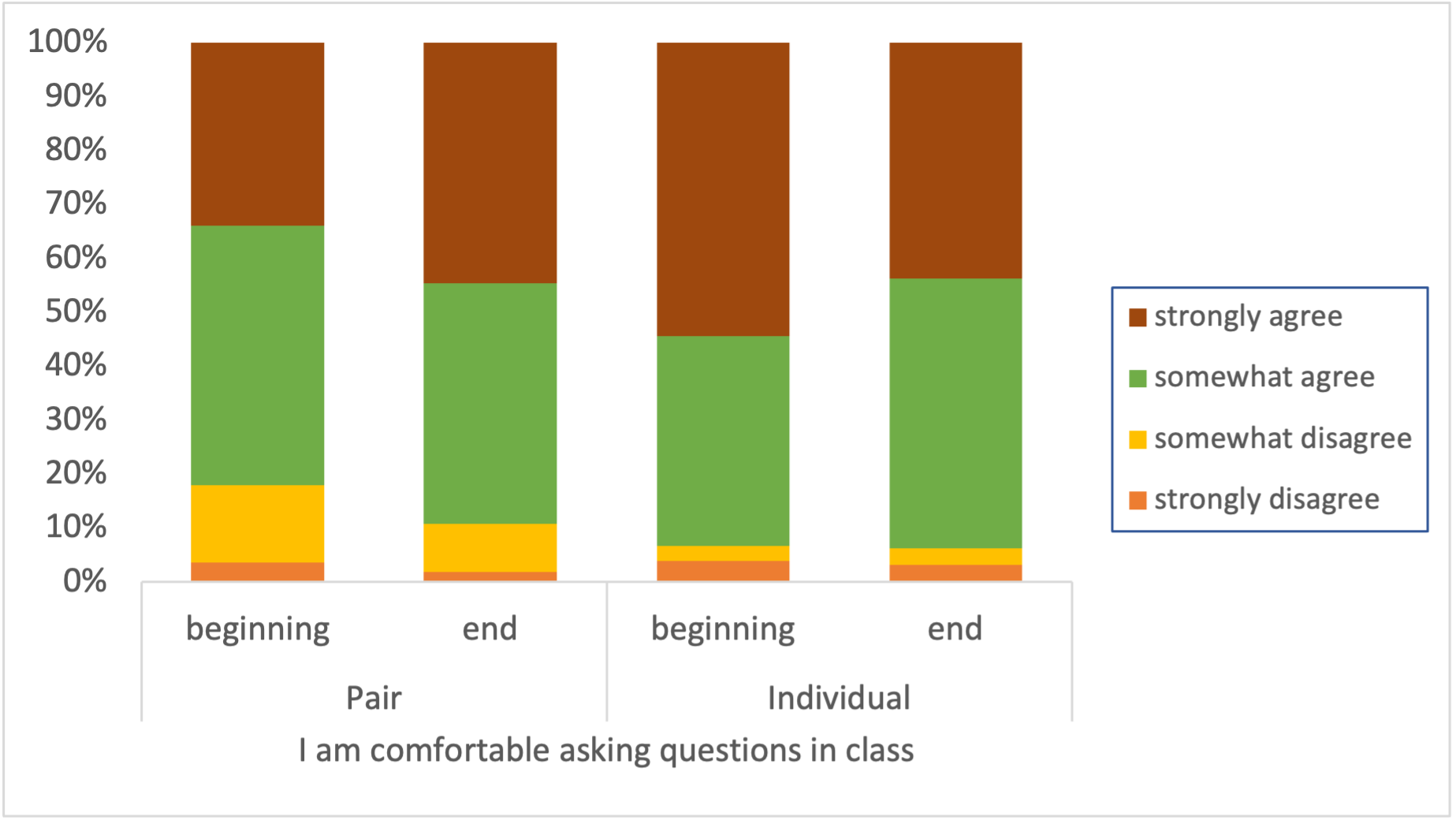}
  \caption{Pair versus Individual Confidence Asking Questions in Class}
  \label{fig:pQuestions}
\end{figure}

Finally, individual students were initially more comfortable asking questions in class than pair programming students (\autoref{fig:pQuestions}). However, by the end of the semester, pair programming students had slightly surpassed the individuals in number of students who ``strongly agreed" that ``I am comfortable asking questions in class" (44.6\% vs. 43.8\%). This suggests that pair programming may positively affect student confidence related to speaking up in class.

\subsection{In Person vs. Remote}
Students in the pair programming group were asked to respond to the statement “my pair programming partner and I always worked remotely” on the end of semester survey. Overall, 34 (59\%) strongly agreed, 21 (36\%) somewhat agreed, 2 (3\%) somewhat disagreed, and 1 (2\%) strongly disagreed. It appears one pair had a student who somewhat agreed with the statement, while the other somewhat disagreed. Ultimately, while the course itself was offered in person, most students chose to pair program remotely. While we did not ask the students why they chose to do so, we suspect it was due to convenience (several students mentioned something similar to struggling with finding a common time to meet) or safety concerns (e.g., to protect against the transmission of Covid-19).

Programming assignments and final scores were compared between those who agreed or strongly agreed to working remotely and those who disagreed or strongly disagreed (i.e., worked in person). An interesting trend emerged (\autoref{tab:remoteOnScores}). The T-tests showed no statistically significant difference between both groups on any assessment, except for assignment 4, in which the group that worked remotely did better (by 2.9 points) than those that worked in person (p = .012). In addition, the actual difference between remote and in person groups narrowed over time, suggesting that those who pair programmed remotely found a routine that worked as the semester progressed. This means that no adverse effects of remote pair programming on student outcomes were observed in this study.


\begin{table}
  \caption{Effect of Pair Programming Remotely on Scores}
  \label{tab:remoteOnScores}
  \begin{tabular}{lrrlr}
\toprule
&Remote&In Person&Test&P-value\\
&Mean (SD)&Mean (SD)&&\\
\midrule
Assignment 1&91.8 (19.0)&100 (0.0)&T-test&0.463\\
Assignment 2&88.3 (28.7)&95.8 (7.2)&T-test&0.655\\
Assignment 3&99.8 (1.1)&100 (0.0)&T-test&0.725\\
Assignment 4&99.6 (1.9)&96.7 (2.9)&T-test&0.012\\
&&&&\\
Final Score&87.9 (9.3)&89.6 (6.9)&T-test&0.768\\
  \bottomrule
\end{tabular}
\end{table}

\subsection{Gender Differences}
Women in the pair programming group scored almost 12\% higher than women in the individual programming group on their final score in the course. While this difference did not achieve statistical significance (p = .091) due to the low sample size of women in the individual (non-pair) programming group, a 12\% difference is practically significant in that it could result in a difference of 2 letter grades (e.g., from a 79 to a 91).

\begin{figure}[h]
  \centering
  \includegraphics[width=\linewidth]{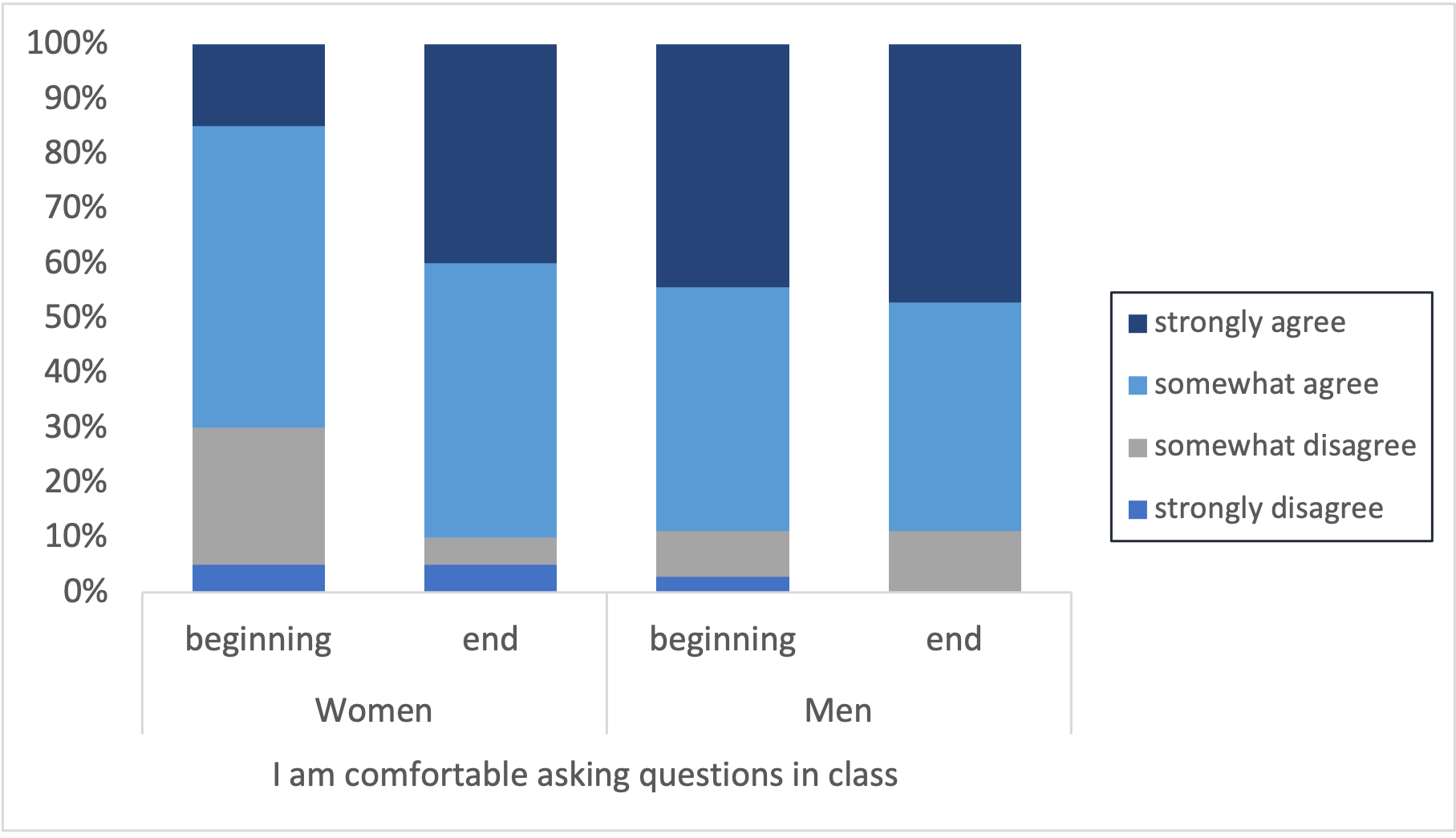}
  \caption{Confidence Asking Questions in Class by Gender}
  \label{fig:questions}
\end{figure}

There were also several gender differences in responses on the end of semester survey from students in the pair programming section. First, women reported an increase in comfort asking questions in class, when compared to men (\autoref{fig:questions}). At the beginning of the semester, only 15\% strongly agreed with the statement, ``I am comfortable asking questions in class". At the end of the semester, 40\% of women students agreed with the statement, and fewer women than men disagreed with the statement. This is important, because it is an indicator of confidence and implies the ability to handle graceful failure in front of peers. By participating in classroom discussion, there is greater opportunity for learning.

\begin{figure}[h]
  \centering
  \includegraphics[width=\linewidth]{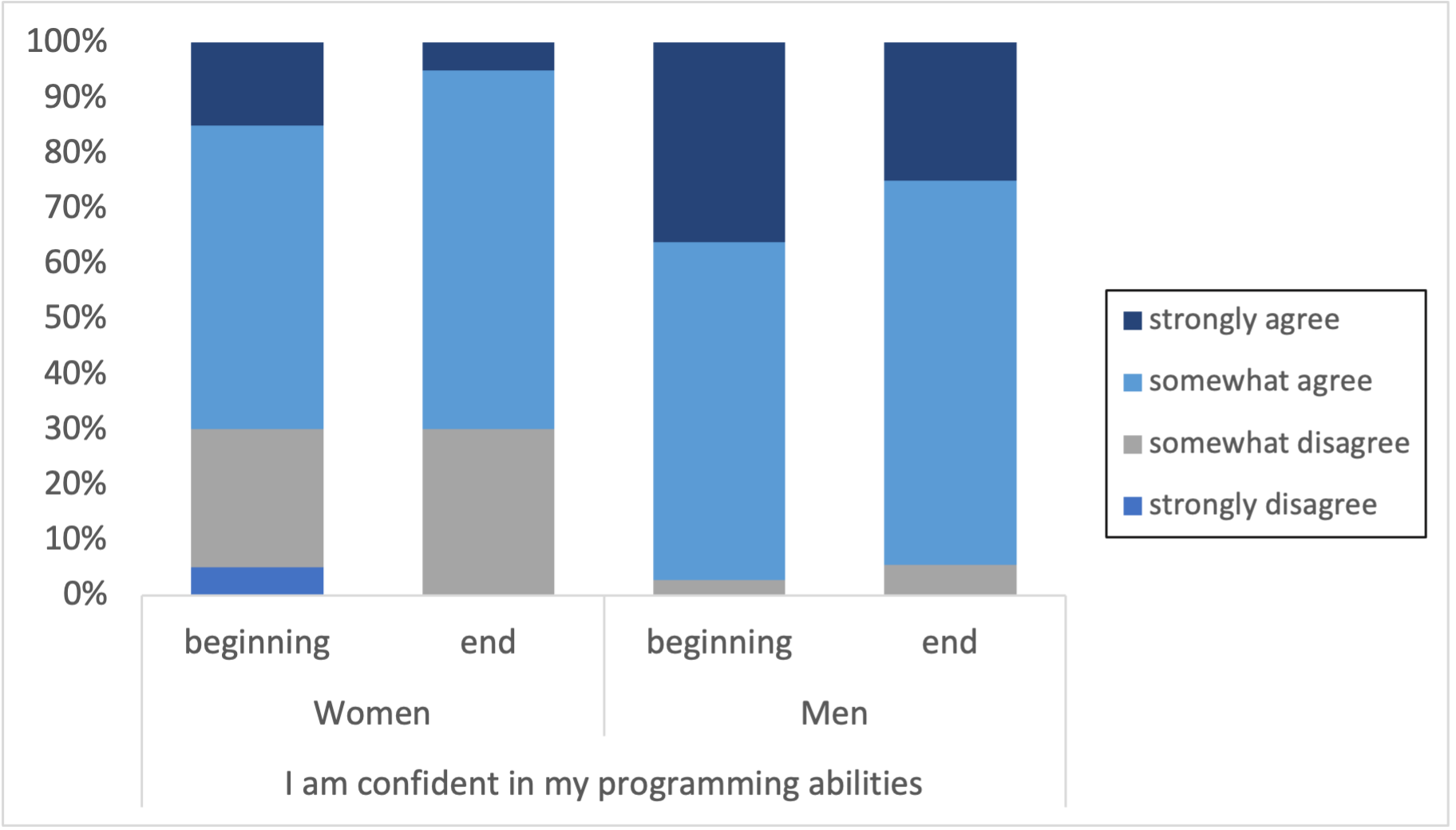}
  \caption{Confidence in Programming Abilities by Gender}
  \label{fig:abilities}
\end{figure}

Both men and women saw a similar decrease in confidence in their programming abilities by the end of the semester (a reduction of 10\% in ``strongly agree", \autoref{fig:abilities}). In contrast to a first year course, where many students are programming for the first time and seeing marked strides in learning a new language, students in a CS-3 course like Data Structures have prior experience, but are then exposed to significantly difficult assignments. The learning curve is much higher. Thus, it is not surprising that some students adjusted their confidence after experiencing the rigor of solving large, real-world problems.

\begin{figure}[h]
  \centering
  \includegraphics[width=\linewidth]{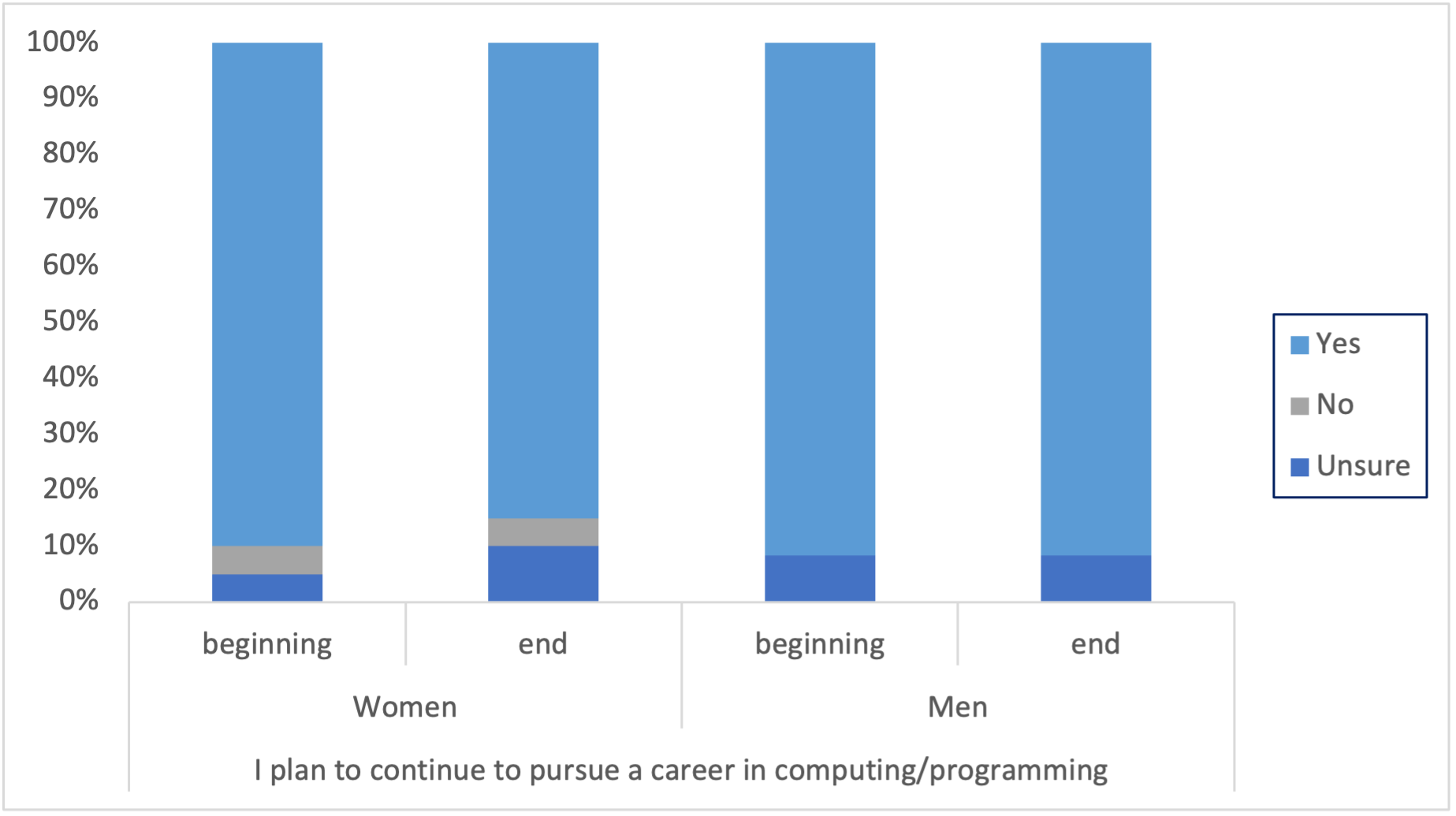}
  \caption{Continue a Career in Computing by Gender}
  \label{fig:career}
\end{figure}

No male student indicated that he had a definitive plan to pursue a career outside of computer science (\autoref{fig:career}). Some women, however, noted definitively that they did not intend to continue work in computer science. A representative comment that explains the sentiment behind taking Data Structures, but not desiring to continue in the field is the following from a woman in the first section: ``I feel as though I am a people person and I would not enjoy sitting behind a computer computing/programming all day. I'd rather sell the product [of] the computation to people, but have the understanding of what it is doing." 

Finally, when asked to respond to the statement “I prefer pair programming to working individually” women had more extreme responses than men. Almost one-quarter (23.8\%) of female students strongly agreed compared to 8.1\% of male students. Similarly, one-third of women (33.3\%) strongly disagreed compared to 18.9\% of men. The differences in responses to this question were marginally significant with p = .052.

\section{Discussion}

Over the semester, we noted several observed benefits from the instructor and TA perspective. First, pair programming promotes social intelligence and teamwork capability, two important skills for programmers today. Pair programming more closely models work in industry, where most students will work in teams. While there were a few groups with issues (mainly related to a failure to properly plan ahead for meetings), pair programming afforded the opportunity for students to communicate and resolve differences. Many students mentioned in the survey enjoying this aspect of pair programming. Particularly as our institution shifted to offering in person courses in the wake of Covid-19, students were eager for interaction and social benefits provided by peer review \cite{celepkolu2018thematic}.

Pair programming may reduce instances of plagiarism and promote academic integrity because a student must 1) decide to cheat and 2) convince the pair programming partner to become complicit. The social stigma against suggesting cheating is a useful barrier. In addition, if a pair programming group is comprised of at least one student who plans and works ahead, then it can reduce procrastination which can lead to cheating out of desperation because of an impending deadline.

We also observed that pair programming reduces the number of questions asked on assignments, suggesting that students successfully use one another as a resource. Each Data Structures course had a separate Piazza page (crowd-sourced communication tool for professors, TAs, and even other students to provide clarification or help to students' questions). Students were encouraged to post there first for a faster response (both sections had a 9-minute average response time at the end of the semester). In the Data Structures 1 (with pair programming) students posted an average of 1.05 questions per student, while students in Data Structures 2 (without pair programming) posted an average of 1.72 questions per student. This suggests another intangible benefit to pair programming: that students can, and will, use their partner as a primary resource for clarifying questions.



\begin{figure*}[hbt!]
\centering
  \includegraphics[height=4.75cm]{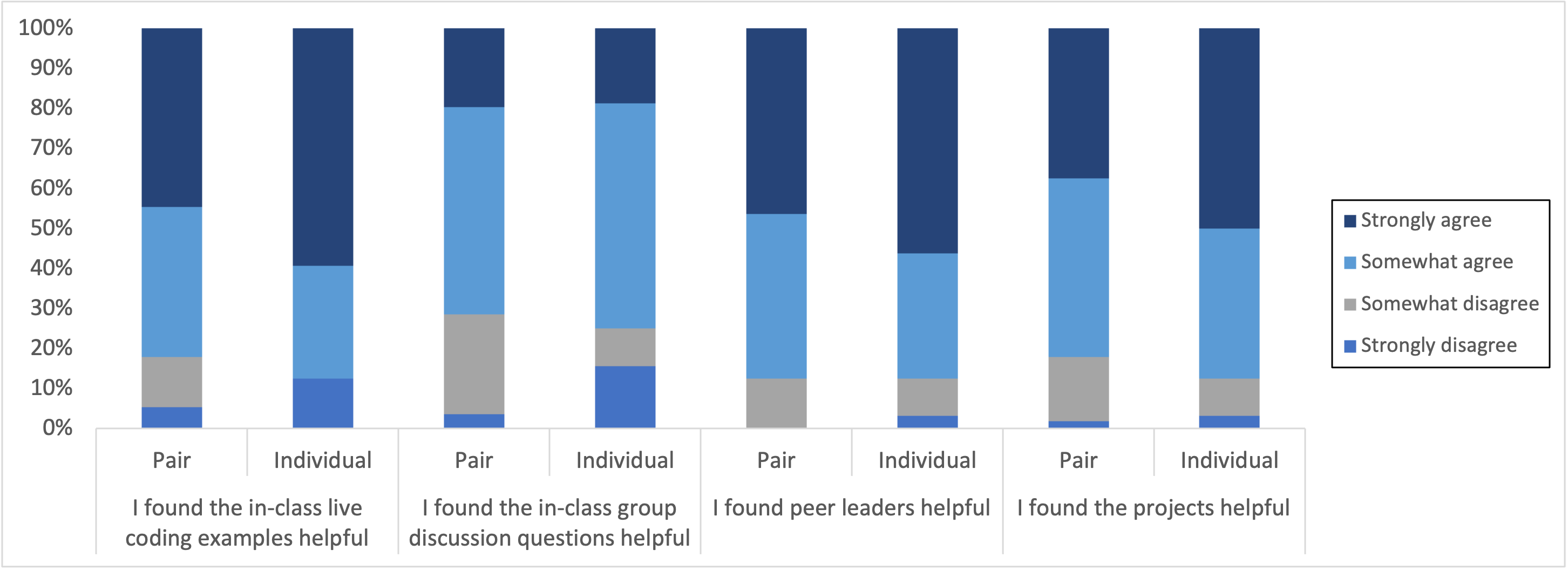}
  \caption{Pair versus Individual Active Learning Technique Preference}
  \label{fig:aL}
\end{figure*}

Finally, several other active learning techniques were utilized in addition to pair programming on assignments in both sections of Data Structures: in-class small group discussion questions modeled after peer instruction \cite{porter2013halving}, peer leaders \cite{bowling2015professionalizing}, and live coding \cite{rubin2013effectiveness}. \autoref{fig:aL} shows how helpful the students found each active learning technique in the post-course survey. The men appeared to prefer peer leaders and live coding, while the women preferred the assignments and live coding examples. In-class group discussion questions appears to be the least helpful to all students. Note that this does not mean the practice is not beneficial to students, as perceived helpfulness is not always correlated to actual helpfulness.

\section{Conclusion and Future Work}

Pair programming provides an advantage over individual coding in overall performance in the classroom, even when pairs collaborate remotely. This was demonstrated by the pair programming groups' better performance on exams and on their final course score. The exam score differences between the two groups appeared to increase over time. Pair programming also demonstrated an advantage on programming assignments. Women in the pair programming group had an almost 12 percentage point advantage over their individually programming counterparts. They also experienced an increase in comfort asking questions in class over the course of the semester to more closely match that of the men. Thus we have found on several fronts that pair programming is one technique that can be utilized to support the goal of increasing the participation of women in STEM professions.

For future work, we intend to 1) delve more deeply into students who do well on paired assignments, but poorly on individual exams. Such students may simply do better on programming projects rather than conceptual questions, or it may be a result of team breakdown. Thus, we further plan to 2) study the optimal makeup of teams, based on gender, prior ability, and working styles (e.g., supporter versus leader or early submitter versus late submitter). We will also study what makes an ideal pair programming partner and analyze the results with a natural language processing algorithm to identify common trends (\textit{aspects} with correlated sentiment) \cite{beasley2021polarity}. In doing so, we will better understand the team dynamics and optimal pairing techniques to allow computer science students to realize the full benefits of pair programming.

\begin{acks}
The authors are grateful to Dr. Jing Wang for her support. This material is based upon work supported by USF STEER, USF CITL, and the National Center for Women \& Information Technology. We would also like to thank the reviewers for their helpful comments and suggestions.
\end{acks}

\bibliographystyle{ACM-Reference-Format}
\bibliography{pair-programming}


\begin{thebibliography}{19}


\ifx \showCODEN    \undefined \def \showCODEN     #1{\unskip}     \fi
\ifx \showDOI      \undefined \def \showDOI       #1{#1}\fi
\ifx \showISBNx    \undefined \def \showISBNx     #1{\unskip}     \fi
\ifx \showISBNxiii \undefined \def \showISBNxiii  #1{\unskip}     \fi
\ifx \showISSN     \undefined \def \showISSN      #1{\unskip}     \fi
\ifx \showLCCN     \undefined \def \showLCCN      #1{\unskip}     \fi
\ifx \shownote     \undefined \def \shownote      #1{#1}          \fi
\ifx \showarticletitle \undefined \def \showarticletitle #1{#1}   \fi
\ifx \showURL      \undefined \def \showURL       {\relax}        \fi
\providecommand\bibfield[2]{#2}
\providecommand\bibinfo[2]{#2}
\providecommand\natexlab[1]{#1}
\providecommand\showeprint[2][]{arXiv:#2}

\bibitem[\protect\citeauthoryear{Adeliyi, Wermelinger, Kear, and
  Rosewell}{Adeliyi et~al\mbox{.}}{2021}]%
        {adeliyi2021investigating}
\bibfield{author}{\bibinfo{person}{Adeola Adeliyi}, \bibinfo{person}{Michel
  Wermelinger}, \bibinfo{person}{Karen Kear}, {and} \bibinfo{person}{Jon
  Rosewell}.} \bibinfo{year}{2021}\natexlab{}.
\newblock \showarticletitle{Investigating Remote Pair Programming In Part-Time
  Distance Education}. In \bibinfo{booktitle}{\emph{United Kingdom and Ireland
  Computing Education Research conference.}} \bibinfo{pages}{1--7}.
\newblock


\bibitem[\protect\citeauthoryear{Baheti, Gehringer, and Stotts}{Baheti
  et~al\mbox{.}}{2002}]%
        {baheti2002exploring}
\bibfield{author}{\bibinfo{person}{Prashant Baheti}, \bibinfo{person}{Edward
  Gehringer}, {and} \bibinfo{person}{David Stotts}.}
  \bibinfo{year}{2002}\natexlab{}.
\newblock \showarticletitle{Exploring the efficacy of distributed pair
  programming}. In \bibinfo{booktitle}{\emph{Conference on Extreme Programming
  and Agile Methods}}. Springer, \bibinfo{pages}{208--220}.
\newblock


\bibitem[\protect\citeauthoryear{Beasley, Piegl, and Rosen}{Beasley
  et~al\mbox{.}}{2021}]%
        {beasley2021polarity}
\bibfield{author}{\bibinfo{person}{Zachariah Beasley}, \bibinfo{person}{Les~A
  Piegl}, {and} \bibinfo{person}{Paul Rosen}.} \bibinfo{year}{2021}\natexlab{}.
\newblock \showarticletitle{Polarity in the Classroom: A Case Study Leveraging
  Peer Sentiment Toward Scalable Assessment}.
\newblock \bibinfo{journal}{\emph{IEEE Transactions on Learning Technologies}}
  (\bibinfo{year}{2021}).
\newblock


\bibitem[\protect\citeauthoryear{Bowling}{Bowling}{2015}]%
        {bowling2015professionalizing}
\bibfield{author}{\bibinfo{person}{Bethany Bowling}.}
  \bibinfo{year}{2015}\natexlab{}.
\newblock \showarticletitle{Professionalizing the role of peer leaders in
  STEM}.
\newblock \bibinfo{journal}{\emph{Journal of STEM Education}}
  \bibinfo{volume}{16}, \bibinfo{number}{2} (\bibinfo{year}{2015}).
\newblock


\bibitem[\protect\citeauthoryear{Bowman, Jarratt, Culver, and Segre}{Bowman
  et~al\mbox{.}}{2019}]%
        {bowman2019prior}
\bibfield{author}{\bibinfo{person}{Nicholas~A Bowman}, \bibinfo{person}{Lindsay
  Jarratt}, \bibinfo{person}{KC Culver}, {and} \bibinfo{person}{Alberto~Maria
  Segre}.} \bibinfo{year}{2019}\natexlab{}.
\newblock \showarticletitle{How Prior Programming Experience Affects Students'
  Pair Programming Experiences and Outcomes}. In
  \bibinfo{booktitle}{\emph{Proceedings of the 2019 ACM Conference on
  Innovation and Technology in Computer Science Education}}.
  \bibinfo{pages}{170--175}.
\newblock


\bibitem[\protect\citeauthoryear{Celepkolu and Boyer}{Celepkolu and
  Boyer}{2018}]%
        {celepkolu2018thematic}
\bibfield{author}{\bibinfo{person}{Mehmet Celepkolu} {and}
  \bibinfo{person}{Kristy~Elizabeth Boyer}.} \bibinfo{year}{2018}\natexlab{}.
\newblock \showarticletitle{Thematic analysis of students' reflections on pair
  programming in cs1}. In \bibinfo{booktitle}{\emph{Proceedings of the 49th ACM
  technical symposium on computer science education}}.
  \bibinfo{pages}{771--776}.
\newblock


\bibitem[\protect\citeauthoryear{Clark and Raker}{Clark and Raker}{2020}]%
        {clark2020peer}
\bibfield{author}{\bibinfo{person}{Aaron Clark} {and}
  \bibinfo{person}{Jeffrey~R Raker}.} \bibinfo{year}{2020}\natexlab{}.
\newblock \showarticletitle{Peer-Leaders' Perceived Roles: An Exploratory Study
  in a Postsecondary Organic Chemistry Course.}
\newblock \bibinfo{journal}{\emph{International Journal of Teaching and
  Learning in Higher Education}} \bibinfo{volume}{32}, \bibinfo{number}{2}
  (\bibinfo{year}{2020}), \bibinfo{pages}{180--189}.
\newblock


\bibitem[\protect\citeauthoryear{Cockburn and Williams}{Cockburn and
  Williams}{2000}]%
        {cockburn2000costs}
\bibfield{author}{\bibinfo{person}{Alistair Cockburn} {and}
  \bibinfo{person}{Laurie Williams}.} \bibinfo{year}{2000}\natexlab{}.
\newblock \showarticletitle{The costs and benefits of pair programming}.
\newblock \bibinfo{journal}{\emph{Extreme programming examined}}
  \bibinfo{volume}{8} (\bibinfo{year}{2000}), \bibinfo{pages}{223--247}.
\newblock


\bibitem[\protect\citeauthoryear{da~Silva~Est{\'a}cio and
  Prikladnicki}{da~Silva~Est{\'a}cio and Prikladnicki}{2015}]%
        {da2015distributed}
\bibfield{author}{\bibinfo{person}{Bernardo~Jos{\'e} da Silva~Est{\'a}cio}
  {and} \bibinfo{person}{Rafael Prikladnicki}.}
  \bibinfo{year}{2015}\natexlab{}.
\newblock \showarticletitle{Distributed pair programming: A systematic
  literature review}.
\newblock \bibinfo{journal}{\emph{Information and Software Technology}}
  \bibinfo{volume}{63} (\bibinfo{year}{2015}), \bibinfo{pages}{1--10}.
\newblock


\bibitem[\protect\citeauthoryear{Hughes, Walshe, Law, and Murphy}{Hughes
  et~al\mbox{.}}{2020}]%
        {hughes2020remote}
\bibfield{author}{\bibinfo{person}{Janet Hughes}, \bibinfo{person}{Ann Walshe},
  \bibinfo{person}{Bobby Law}, {and} \bibinfo{person}{Brendan Murphy}.}
  \bibinfo{year}{2020}\natexlab{}.
\newblock \showarticletitle{Remote pair programming}. In
  \bibinfo{booktitle}{\emph{12th International Conference on Computer Supported
  Education}}. SciTePress, \bibinfo{pages}{476--483}.
\newblock


\bibitem[\protect\citeauthoryear{Kuttal, Gerstner, and Bejarano}{Kuttal
  et~al\mbox{.}}{2019}]%
        {kuttal2019remote}
\bibfield{author}{\bibinfo{person}{Sandeep~Kaur Kuttal}, \bibinfo{person}{Kevin
  Gerstner}, {and} \bibinfo{person}{Alexandra Bejarano}.}
  \bibinfo{year}{2019}\natexlab{}.
\newblock \showarticletitle{Remote pair programming in online cs education:
  Investigating through a gender lens}. In \bibinfo{booktitle}{\emph{2019 IEEE
  Symposium on Visual Languages and Human-Centric Computing (VL/HCC)}}. IEEE,
  \bibinfo{pages}{75--85}.
\newblock


\bibitem[\protect\citeauthoryear{McDowell, Werner, Bullock, and
  Fernald}{McDowell et~al\mbox{.}}{2002}]%
        {mcdowell2002effects}
\bibfield{author}{\bibinfo{person}{Charlie McDowell}, \bibinfo{person}{Linda
  Werner}, \bibinfo{person}{Heather Bullock}, {and} \bibinfo{person}{Julian
  Fernald}.} \bibinfo{year}{2002}\natexlab{}.
\newblock \showarticletitle{The effects of pair-programming on performance in
  an introductory programming course}. In \bibinfo{booktitle}{\emph{Proceedings
  of the 33rd SIGCSE technical symposium on Computer science education}}.
  \bibinfo{pages}{38--42}.
\newblock


\bibitem[\protect\citeauthoryear{Porter, Bailey~Lee, and Simon}{Porter
  et~al\mbox{.}}{2013}]%
        {porter2013halving}
\bibfield{author}{\bibinfo{person}{Leo Porter}, \bibinfo{person}{Cynthia
  Bailey~Lee}, {and} \bibinfo{person}{Beth Simon}.}
  \bibinfo{year}{2013}\natexlab{}.
\newblock \showarticletitle{Halving fail rates using peer instruction: a study
  of four computer science courses}. In \bibinfo{booktitle}{\emph{Proceeding of
  the 44th ACM technical symposium on Computer science education}}.
  \bibinfo{pages}{177--182}.
\newblock


\bibitem[\protect\citeauthoryear{ROBE and KUTTAL}{ROBE and KUTTAL}{2021}]%
        {robe2021designing}
\bibfield{author}{\bibinfo{person}{PETER ROBE} {and}
  \bibinfo{person}{SANDEEP~KAUR KUTTAL}.} \bibinfo{year}{2021}\natexlab{}.
\newblock \showarticletitle{Designing PairBuddy--A Conversational Agent for
  Pair Programming}.
\newblock  (\bibinfo{year}{2021}).
\newblock


\bibitem[\protect\citeauthoryear{Rubin}{Rubin}{2013}]%
        {rubin2013effectiveness}
\bibfield{author}{\bibinfo{person}{Marc~J Rubin}.}
  \bibinfo{year}{2013}\natexlab{}.
\newblock \showarticletitle{The effectiveness of live-coding to teach
  introductory programming}. In \bibinfo{booktitle}{\emph{Proceeding of the
  44th ACM technical symposium on Computer science education}}.
  \bibinfo{pages}{651--656}.
\newblock


\bibitem[\protect\citeauthoryear{Umapathy and Ritzhaupt}{Umapathy and
  Ritzhaupt}{2017}]%
        {umapathy2017meta}
\bibfield{author}{\bibinfo{person}{Karthikeyan Umapathy} {and}
  \bibinfo{person}{Albert~D Ritzhaupt}.} \bibinfo{year}{2017}\natexlab{}.
\newblock \showarticletitle{A meta-analysis of pair-programming in computer
  programming courses: Implications for educational practice}.
\newblock \bibinfo{journal}{\emph{ACM Transactions on Computing Education
  (TOCE)}} \bibinfo{volume}{17}, \bibinfo{number}{4} (\bibinfo{year}{2017}),
  \bibinfo{pages}{1--13}.
\newblock


\bibitem[\protect\citeauthoryear{Williams and Upchurch}{Williams and
  Upchurch}{2001}]%
        {williams2001support}
\bibfield{author}{\bibinfo{person}{Laurie Williams} {and}
  \bibinfo{person}{Richard~L Upchurch}.} \bibinfo{year}{2001}\natexlab{}.
\newblock \showarticletitle{In support of student pair-programming}.
\newblock \bibinfo{journal}{\emph{ACM SIGCSE Bulletin}} \bibinfo{volume}{33},
  \bibinfo{number}{1} (\bibinfo{year}{2001}), \bibinfo{pages}{327--331}.
\newblock


\bibitem[\protect\citeauthoryear{Williams and Kessler}{Williams and
  Kessler}{2000}]%
        {williams2000all}
\bibfield{author}{\bibinfo{person}{Laurie~A Williams} {and}
  \bibinfo{person}{Robert~R Kessler}.} \bibinfo{year}{2000}\natexlab{}.
\newblock \showarticletitle{All I really need to know about pair programming I
  learned in kindergarten}.
\newblock \bibinfo{journal}{\emph{Commun. ACM}} \bibinfo{volume}{43},
  \bibinfo{number}{5} (\bibinfo{year}{2000}), \bibinfo{pages}{108--114}.
\newblock


\bibitem[\protect\citeauthoryear{Ying, Pezzullo, Ahmed, Crompton, Blanchard,
  and Boyer}{Ying et~al\mbox{.}}{2019}]%
        {ying2019their}
\bibfield{author}{\bibinfo{person}{Kimberly~Michelle Ying},
  \bibinfo{person}{Lydia~G Pezzullo}, \bibinfo{person}{Mohona Ahmed},
  \bibinfo{person}{Kassandra Crompton}, \bibinfo{person}{Jeremiah Blanchard},
  {and} \bibinfo{person}{Kristy~Elizabeth Boyer}.}
  \bibinfo{year}{2019}\natexlab{}.
\newblock \showarticletitle{In their own words: Gender differences in student
  perceptions of pair programming}. In \bibinfo{booktitle}{\emph{Proceedings of
  the 50th ACM Technical Symposium on Computer Science Education}}.
  \bibinfo{pages}{1053--1059}.
\newblock


\end{thebibliography}

\appendix

\end{document}